\documentclass[12pt]{spieman}  
\usepackage{amsmath,amsfonts,amssymb}
\usepackage{graphicx}
\usepackage{setspace}
\usepackage{tocloft}
\usepackage{lineno}
\usepackage{color,soul}
\linenumbers
\title{Understanding the Evolution of Black Hole Accretion and Dust out to $z=4$ with a Deep Imaging Extragalactic Survey with {\it PRIMA}}

\author[a,*]{Andreas L. Faisst}
\author[b]{Chian-Chou Chen}
\author[c]{Laure Ciesla}
\author[d]{Carlotta Gruppioni}
\affil[a]{Caltech/IPAC, 1200 E. California Blvd. Pasadena, CA 91125, USA}
\affil[b]{Academia Sinica Institute of Astronomy and Astrophysics (ASIAA), No. 1, Sec. 4, Roosevelt Road, Taipei 106319, Taiwan}
\affil[c]{Aix Marseille Univ, CNRS, CNES, LAM, Marseille, France}
\affil[d]{Istituto Nazionale di Astrofisica - Osservatorio di Astrofisica e Scienza dello Spazio, via Gobetti 93/3, Bologna, Italy, I-40129}

\DeclareRobustCommand{\ion}[2]{%
\relax\ifmmode
\ifx\testbx\f@series
{\mathbf{#1\,\mathsc{#2}}}\else
{\mathrm{#1\,\mathsc{#2}}}\fi
\else\textup{#1\,{\mdseries\textsc{#2}}}%
\fi}

\def \h2{{\rm H_{2}}}

\def \LIR{L_{{\rm IR}}}

\def \dn4000{D_{{\rm n}}(4000) }

\cftpagenumbersoff{figure}
\cftpagenumbersoff{table} 
\begin{document}
\nolinenumbers
\maketitle

\begin{abstract}

The cosmic evolution of obscured star formation, dust properties and production mechanisms, and the prevalence of dust-obscured AGN out to high redshifts are currently some of the hot topics in astrophysics. While much progress has been made in the early days with {\it Spitzer} and {\it Herschel}, these facilities have not reached the necessary depths to observe the mid-IR light of high-redshift ($z>3$) galaxies. Recently, the {\it James Webb Space Telescope} (JWST) has filled in the blue side of the rest-frame mid-IR. The {\it Atacama Large (Sub)Millimeter Array} (ALMA), on the other hand, provides excellent sensitivity in the far-IR regime, allowing the study of dust and gas properties at high redshifts. Filling the wavelength gap between JWST and ALMA is crucial to progress our understanding of early galaxy evolution -- and this will be an important goal in the next decades. The {\it Probe far-IR Mission for Astrophysics} (PRIMA), with sensitive imaging and spectroscopic capabilities at $24-240\,{\rm \mu m}$ and currently in Phase A study, will achieve this and provide insights into early galaxy evolution, Black Hole growth, and dust production mechanisms. Here we present PRIDES, a possible deep and wide-area survey over $1.6\,{\rm deg^2}$ of the COSMOS field with {\it PRIMA} to study these science cases. 

\end{abstract}


{\noindent \footnotesize\textbf{*} Corresponding author email: \linkable{afaisst@caltech.edu}}

\begin{spacing}{2}   

\section{Introduction}
\label{sec:intro}

\begin{figure*}
\begin{center}
\includegraphics[width=\textwidth]{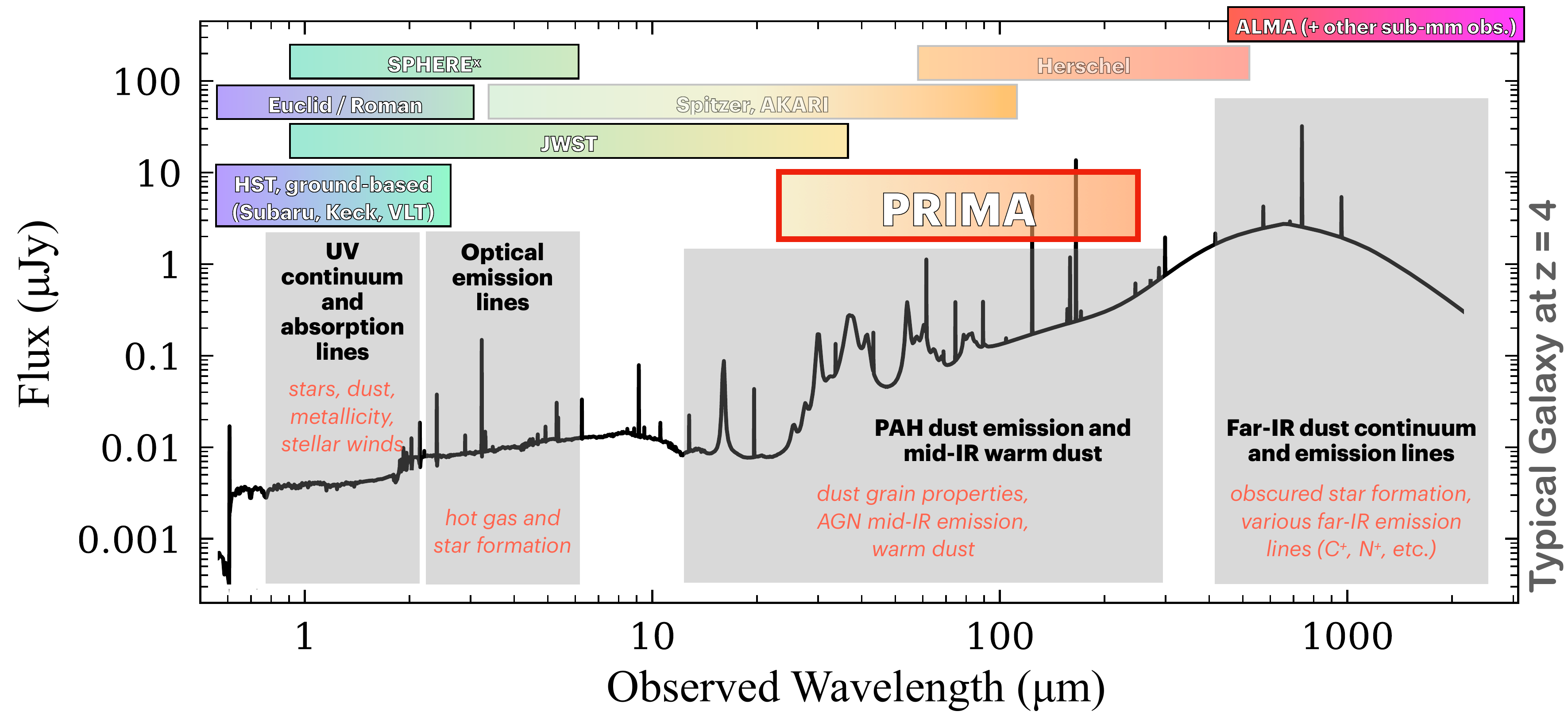}\vspace{-3mm}
\end{center}
\caption 
{\label{fig:coverage}
The wavelength coverage of various past ({\it Spitzer} and {\it Herschel}, grayed out), current, and future ({\it Euclid}, {\it Roman}, and SPHEREx) facilities compared to {\it PRIMA}. Various interesting wavelength regions are indicated. {\it PRIMA} optimally complements these surveys and provides observations in a niche wavelength region that is crucial to progress our understanding of various studies including the evolution of galaxies, black holes, and dust. The model (generated with the \texttt{FSPS} Python package\cite{conroy09,conroy10,johnson24} shows a typical star-forming galaxy at $z\sim4$ as example.} 
\end{figure*}

A multi-wavelength picture of galaxies is crucial to understand their physics. The measurement of the rest-frame ultra-violet (UV) light provides indicators of strong ionizing sources, binary star populations, stellar metallicities and dust absorption, the physics of reionization, dust unobscured star formation, and unobscured feeding supermassive black holes (so called Active Galactic Nuclei, AGN).
On the other hand, rest-frame optical light provides insights into the gas and nebular properties such as ionization rates, gas-phase metal enrichment, and the kinematic structure of the ionized gas. 
On the other end of the spectrum, the far-infrared (far-IR) light close to the peak of far-IR emission ($\sim100\,{\rm \mu m}$ rest-frame) is sensitive to dust mass, dust-obscured star formation, and the warm interstellar medium (ISM) through the measurement of far-IR lines (such as $C^+$~at $158\,{\rm \mu m}$).
As shown in Figure~\ref{fig:coverage}, ground-based facilities as well as the {\it Hubble Space Telescope} (HST) cover the rest-frame UV part of high-redshift galaxies (here $z=5$). The {\it Atacama Large (Sub)Millimeter Array} (ALMA) covers the far-IR part of the spectrum at unprecedented sensitivity. The recently launched {\it James Webb Space Telescope} (JWST) covers successfully the rest-frame optical range with imaging and spectroscopy with its NIRCam and MIRI instruments. 

So far, the {\it Spitzer} and {\it Herschel} space telescopes were covering the mid-IR part of the spectrum, thereby including important indicators of dust and AGN activity. Both facilities are now retired, but their legacy data products are still relevant to science. For example, stacking of {\it Spitzer} and {\it Herschel} imaging provides average dust-obscured star formation rates at cosmic noon ($z\sim2$) and can even constrain the average far-IR spectral energy distribution (SED) of galaxies at $z=5$\cite{bethermin20,schreiber15}.
However, individual galaxies at high redshifts are far too faint to be detected in multiple photometric bands without stacking by these facilities. This therefore result in a serious wavelength gap, which needs to be filled to understand the physics and the interplay of galaxy evolution, star formation rates (SFR), AGN, and dust production at early times beyond cosmic noon.

Studying the SED of galaxies at IR wavelengths is crucial for understanding galaxy formation and evolution: the total IR luminosity probes dust-obscured star formation, and various silicate and polycyclic aromatic hydrocarbon (PAH) spectral features allow us to study dust grain properties and the prevalence of dust-obscured AGN. Recently, wide-field targeted and blind-field surveys with ALMA, such as {\it ALPINE} ($z=4-6$)\cite{lefevre20,bethermin20,faisst20}, {\it REBELS} ($z > 6.5$)\cite{bouwens22}, the (Ex-)MORA ALMA survey ($z = 3 – 6$)\cite{casey21,long24}, have shown considerable dust obscuration even in the early Universe at $z > 3$\cite{gruppioni20,fudamoto20,fudamoto21,talia21,inami22,gentile24}, which was surprising given that dust production is a function of time. New efforts such as the upcoming {\it CHAMPS} large ALMA program (Faisst et al. {\it in prep}) on the COSMOS field\cite{scoville07} provide deep blind-surveys of dusty sources in the early Universe, emphasizing the importance of such studies.
More than 70\% of the light of galaxies at $z\sim3$ is reprocessed by dust ({\it i.e.} emitted in the IR) and about 20\% of galaxies are completely missed in the UV\cite{bouwens09,franco18,wang19}. Some of these ``optical dark'' galaxies are detected in recent JWST imaging, but most are too faint to derive robust redshifts. Unfortunately, all the above ALMA surveys cover only a small area on sky and provide only one data point at IR wavelengths at various redshifts. With currently no constraints on the shape of IR SEDs of $z>3$ galaxies (except some constraints from stacking\cite{schreiber17,schreiber18,bethermin15,bethermin20} in the mid-IR range), we are completely ignorant about dust temperature, the general dust properties, and obscured AGN contribution. This results in significant uncertainty in almost any properties derived from the IR SED\cite{faisst17,faisst20}. Large statistical samples are furthermore needed to study the dust properties and production mechanisms across cosmic time. On another front, JWST recently confirmed AGN via rest-frame UV and optical emission lines in early galaxies out to $z\sim8$\cite{ding22,larson23,matthee24,harikane23,akins24}, indicative of accreting massive black holes which are little dust obscured. Dust-obscured AGN remain hidden at UV/optical wavelengths and even in the X-rays\cite{carroll23}, and mid-IR observations are crucial for their identification. AGN likely have a significant impact on galaxy evolution, as they are thought to be coupled with the quenching of star formation\cite{dubois13}.
These studies lead to several questions: {\it (i)} What is the fraction of obscured star formation across cosmic time? {\it (ii)} How is dust produced and how do its properties change over time? {\it (iii)} What is the prevalence of dust-obscured AGN (missed in rest-UV/optical) and how do they shape galaxy evolution at early stages? {\it (iv)} How do the above parameters change as a function of location on the main sequence and in different environments across cosmic time?

It is clear that for a successful and comprehensive study of early galaxy evolution we need to fill in the mid-IR wavelength gap. The {\it Probe Far-Infrared Mission for Astrophysics} (PRIMA; Glenn et al. {\it in prep.}, this volume), a cryogenically cooled $1.8\,{\rm m}$ diameter space telescope that is currently in Phase A study, is ideally suited for this.
{\it PRIMA} covers a wavelength range of $24-235\,{\rm \mu m}$ with imaging and spectroscopy over a wide field of view, which allows significantly faster mapping speeds than previous and current far-IR telescopes. The sensitivity is significantly improved compared to {\it Spitzer} and {\it Herschel}, capable of detecting the mid-IR emission of galaxies beyond $z=4$. It is however to note that confusion will be one of the main challenges for {\it PRIMA}, which can be remedied somewhat by using state-of-the-art deblending algorithms such as prior-based forced photometry methods. It is planned that {\it PRIMA} will be made available through a General Observer (GO) program offering 75\% of the mission time over 5 years. For a collection of GO science cases, we refer to the {\it PRIMA} General Observer Science Book \cite{moullet23}.

In this work, we introduce a possible survey using {\it PRIMA} observations over a $1.6\,{\rm deg^2}$ area on the COSMOS field, which will allow constraining the mid-IR range of several $10,000$ galaxies out to $z=4$ combined with exquisite ancillary data at UV, optical, and sub-mm wavelength. With this survey, we will be able to answer some of the most relevant questions in current astrophysics, including:
{\it (i)} What is the redshift distribution of optical dark galaxies? 
{\it (ii)} What is the contribution of dust-obscured star formation and how do the dust properties evolve across redshifts?
{\it (iii)} What is the prevalence of dust-obscured AGN in the early Universe?

In Section~\ref{sec:need}, we iterate the need for a mid-IR facility such as {\it PRIMA} to study the rise of dust and metals, obscured star formation, and AGN across cosmic time. In Section~\ref{sec:prides}, we introduce the {\it PRIMA} Imaging Deep Extragalactic Survey (PRIDES) which is built to answer the above questions. Finally, we conclude in Section~\ref{sec:conclusions}.
Throughout this work, we assume a $\Lambda$CDM cosmology with $H_0 = 70\,{\rm km\,s^{-1}\,Mpc^{-1}}$, $\Omega_\Lambda = 0.7$, and $\Omega_{\rm m} = 0.3$ and magnitudes are given in the AB system\cite{oke74}. The total IR luminosity ($L_{\rm IR}$) refers to the integral of the IR SED at rest-frame $8-1000\,{\rm \mu m}$. The stellar masses and SFRs are normalized to a Chabrier\cite{chabrier03} initial mass function (IMF).


\section{The Need for {\it PRIMA}}\label{sec:need}

{\it PRIMA} (PI: Jason Glenn; Glenn et al., {\it in prep.}, this volume) is a new space-based facility that was selected for a Phase A study in October 2024. It is a cryogenically cooled 1.8 meter diameter space telescope covering the wavelength range from $24-235\,{\rm \mu m}$ with imaging and spectroscopy. The imaging component (PRIMAger; Ciesla et al. {\it in prep.}, this volume) consists of a Hyperspectral imager split in two arrays (covering the wavelength range $24-84\,{\rm \mu m}$ in linear variable filters [LVF] at a resolution of $R=10$, equivalent to $12$ filters) and a Polarimetric imager (covering four bands at 91, 125, 165, and $232\,{\rm \mu m}$). The resolution Point Spread Function (PSF) FWHM varies from $4^{\prime\prime}$ to $24^{\prime\prime}$ and the pixel size varies from $4^{\prime\prime}$ to $24^{\prime\prime}$ from blue to red (generally resulting in a slightly oversampled PSF). LVFs allow for an efficient observing strategy; the wavelength in an exposure varies along one axis of the field of view. By stepping along that axis across the sky, sources fall on different pixels. From this, an image at different wavelengths can be reconstructed, mapping the full wavelength range across the sky.
In addition to the imaging capabilities, {\it PRIMA} offers a spectral model (FIRESS; Bradford et al. {\it in prep.}, this volume). Four gratings (24 spatial $\times$ 84 spectral pixels) covering $24-43\,{\rm \mu m}$, $42-76\,{\rm \mu m}$, $76-134\,{\rm \mu m}$, and $130-235\,{\rm \mu m}$ provide spectra at a resolution of $R\sim100$. In addition, there is the possibility of a high-resolution mode with a Fourier transform module to reach $R\sim 2000-4400$ across the wavelength range.

The field of view of the PRIMAger ranges from $3.8^{\prime}\times3^{\prime}$ to $5^{\prime}\times4.5^{\prime}$ and is thus larger than the field of view of JWST (NIRCam and importantly MIRI). {\it PRIMA} is therefore an excellent facility for surveys. Specifically, its mapping speed (required time to survey a given area to a given depth) is $10-100\times$ faster than JWST/MIRI, $100\times$ faster than {\it Spitzer}, $10^{4-5}\times$ faster than {\it Herschel} PACS and SPIRE, and $10^{5-7}\times$ faster than SOPHIA. As shown in Figure~\ref{fig:coverage}, {\it PRIMA} excellently fills in the current wavelength gap between JWST and ALMA.

PRIMA's core science program addresses three themes:
{\it (i)} the origin of planets and their atmospheres,
{\it (ii)} the co-evolution of galaxies and supermassive black holes since cosmic noon, and
{\it (iii)} the buildup of dust and metals across cosmic time.
The survey presented in this work (Section~\ref{sec:prides}) will focus on the latter two topics. In addition to these topics, the survey will provide a legacy dataset that can be used to carry out a wealth of other science.

Data from past IR facilities such as {\it Spitzer} and {\it Herschel}, which are both retired, are not deep enough (see Section~\ref{sec:need}) to study typical main-sequence galaxies in the early Universe except through stacking \cite{bethermin15,bethermin20,schreiber15}. Furthermore, at these redshifts, ALMA covers wavelengths redward of the IR SED ($\lambda\sim80-100\,{\rm \mu m}$, depending on dust temperature and opacity) as its transmission drops significantly at higher frequencies covering the mid-IR. Hence it is significantly limited in constraining the IR SEDs of high-redshift galaxies. On the other hand, JWST does not go red enough with MIRI. The reduced sensitivity of MIRI at $>20\,{\rm \mu m}$ exacerbate its limits. In both cases, as mentioned above, the small FoV / beam size makes large-area surveys prohibitively expensive.
{\it PRIMA} therefore provides a natural synergy with these observatories by optimally complementing them. Furthermore, the newly launched {\it Euclid} space telescope and future observatories such as {\it SPHEREx}\cite{crill20} and {\it Roman} will provide large-area\footnote{In fact {\it all-sky} in the case of {\it SPHEREx}.} observations at near-IR wavelengths. Again, {\it PRIMA} is a timely complement to these new surveys by the time of its launch and over the next decade.


\section{A Possible Survey: PRIDES -- The {\it PRIMA} imaging Deep Extragalactic Survey}\label{sec:prides}

To answer the questions posed in Section~\ref{sec:intro}, a wide-field survey at mid-IR wavelength is crucial. The fast mapping capabilities, sensitivities, and wavelength coverage of {\it PRIMA} makes it ideally suited for studying the mid-IR light especially at high redshifts ($z>3$), where so far little IR data exists. As mentioned above, {\it PRIMA} directly complements existing observatories (retired, current, and future) and therefore provides {\it the} crucial puzzle piece to push forward the understanding of high-redshift galaxies and their connection to the present galaxies.

In order to do so, samples need to cover various stages of galaxy formation, including different spatial environments. For this, surveying a large and contiguous area on the order of $\sim2\,{\rm deg^2}$ is necessary. {\it PRIMA} is perfectly suited to fill in the gap in wavelength and sensitivity of current facilities, and it is expected that {\it PRIMA} will lead to breakthroughs in the study of dust and AGN of galaxies in the early Universe.

\begin{figure*}
\begin{center}
\includegraphics[width=1\textwidth]{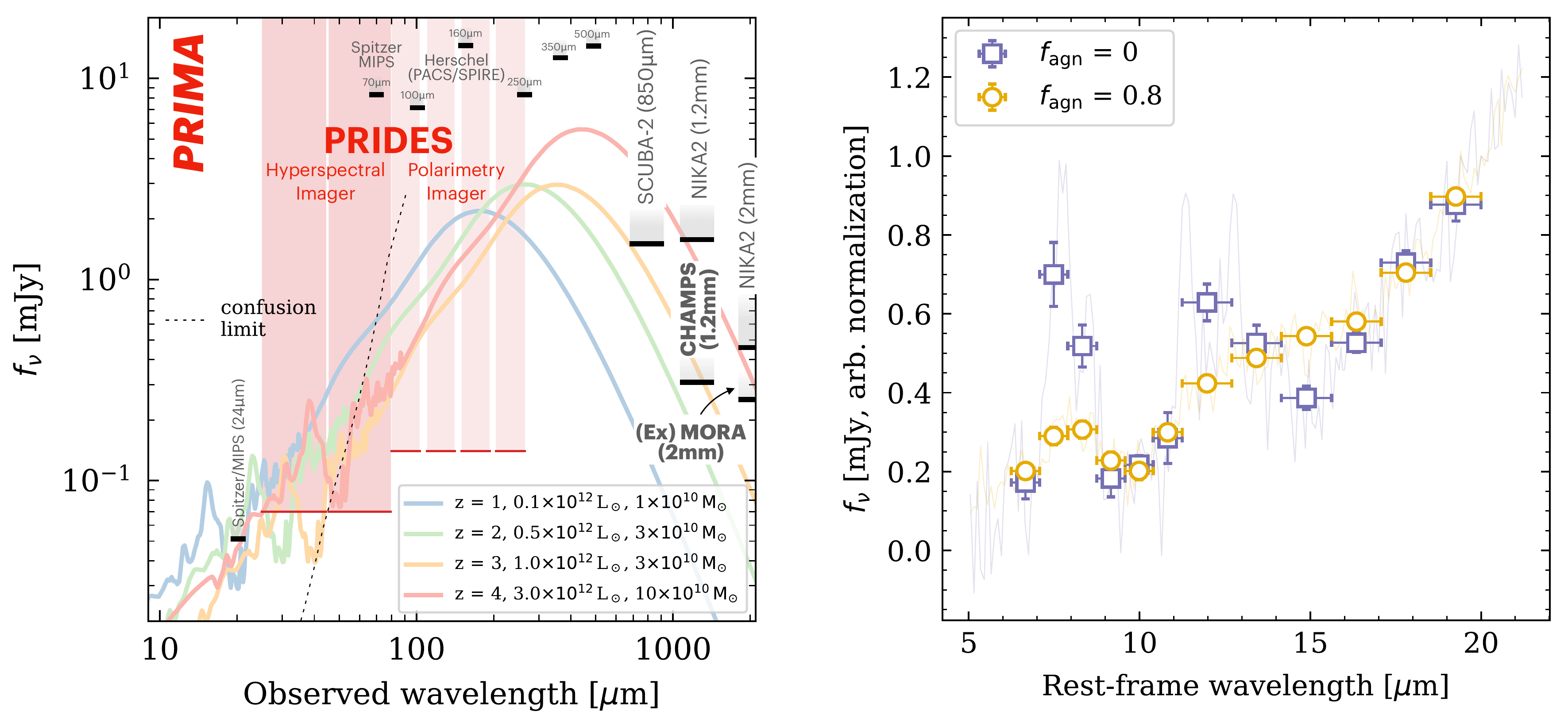}
\vspace{-5mm}
\end{center}
\caption 
{
{\it Left:} Survey depth of the {\it PRIMA} Imaging Deep Extragalactic Survey (PRIDES) to survey a $1.6\,{\rm deg^2}$ area at $70\,{\rm \mu Jy}$ (total time approximately 200 hours). Several different SEDs (from Kirkpatrick et al. 2012\cite{kirkpatrick12}) are shown for different IR luminosities and corresponding stellar masses at $z=1-4$. Also shown is the survey limit of the ALMA programs CHAMPS (Faisst et al., in prep.) and (Ex)MORA\cite{casey21,long24} as well as single-dish surveys SCUBA-2\cite{simpson19} and NIKA2\cite{bing23}. The limits of {\it Spitzer} and {\it Herschel} over the COSMOS field are also indicated\cite{Jin18,frayer09}. (Note that {\it Spitzer}/MIPS $160\,{\rm \mu m}$ limits, $65\,{\rm mJy}$ are not indicated for clarity.) Existing JWST data are significantly deeper ($<10^6\,{\rm \mu m}$) and not shown here for clarity. All limits are $5\sigma$.
{\it Right:} Simulated observations with {\it PRIMA} in Band 1 of a $z=3$ galaxy at $10^{12}\,{\rm L_\odot}$ for 0\% and 80\% AGN contribution\cite{kirkpatrick12,kirkpatrick15} to visualize the measurable changes in the silicate spectral features depending on the hardness of radiation from AGN.
\label{fig:sed}}
\end{figure*}

\subsection{PRIDES -- Science Goals and Synergies}\label{sec:sciencegoals}

Here, we introduce the {\it PRIMA} Imaging Deep Extragalactic Survey (PRIDES) -- a possible GO survey program at about 200 hours designed to observe the rest-frame mid-IR emission of typical galaxies out to $z = 4$ over an area of $1.6\,{\rm deg^2}$. This area is large enough to cover various galaxy properties and environmental regimes and will therefore provide a unique dataset to explore galaxy evolution. PRIDES makes use of PRIMAger in the wavelength range of $25-80\,{\rm \mu m}$ at an $R\sim10$ (see Section~\ref{sec:need}) to a depth of $\sim70\,{\rm \mu Jy}$ and is driven by following three main science goals:
\begin{itemize}
    \item Measure redshifts and SED shapes of hundreds of optical dark galaxies, so far only detected with ALMA and the reddest JWST bands with NIRCam and MIRI, by adding important mid-IR coverage with {\it PRIMA}. Multiple photometric points in the mid-IR blue-ward of the far-IR peak ($\sim100\,{\rm \mu m}$ rest-frame) will put strong constraints on the redshift of such dust-obscured sources, especially in regions where no JWST/MIRI observations exist\cite{gruppioni20}. This will be crucial to trace the cosmic dust-obscured star-formation density at $z>3$, which currently is uncertain because of unreliable distance measurements\cite{gruppioni20,talia21}.
    \item Measure the obscured star formation and dust properties via the abundance of PAH emission and silicate absorption/emission features for the first and most complete census of dust build-up and star formation of the galaxy population at $z>3$. The large survey volume will enable the study of a complete sample in terms of galaxy properties and large-scale structure environment at stellar masses of $>3\times 10^{10}\,{\rm M_\odot}$. The comparison of the derived PAH measurements (measures of the dust grain size distribution) using {\it PRIMA}'s multi-band resolution with measurements of metallicity and ionization hardness from ancillary rest-frame optical spectroscopic data will provide the first statistical study of the relation between dust, metallicity, and ISM conditions in high-redshift galaxies.
    \item Study the prevalence of dust-obscured AGN in the early Universe by measurements of spectral features (such as PAH emission, as PAHs are being disintegrated and removed by the hard radiation of AGN) and steepness of the mid-IR slope to derive the first unbiased census of dust-obscured AGN in typical $z>3$ galaxies. Dust-obscured, Compton-thick AGN will be missed by rest-frame UV/optical emission line indicators as used by JWST (and even in X-rays) and {\it PRIMA} will be effective of finding these and characterize their likely significant impact on the evolution of galaxies at early times.
\end{itemize}

\begin{table}[t!]
\begin{center}       
\begin{tabular}{|c|c|c|c|c|} 
\hline
\rule[-1ex]{0pt}{3.5ex}  Redshift & $\LIR$ & SFR & $M_{*}$ & Expected number  \\
\rule[-1ex]{0pt}{3.5ex}    & $\rm L_\odot$ & $\rm M_\odot\,yr^{-1}$ & $\rm M_\odot$ &   \\
\hline\hline
\rule[-1ex]{0pt}{3.5ex}  $z\sim 1$ & $>10^{11}$ & $>10$ & $>10^{10}$ & $\sim 18,000$  \\
\rule[-1ex]{0pt}{3.5ex}  $z\sim 2$ & $>5\times 10^{11}$ & $>50$ & $>3\times 10^{10}$ & $\sim 4,000$  \\
\rule[-1ex]{0pt}{3.5ex}  $z\sim 3$ & $>10^{12}$ & $>100$ & $>3\times 10^{10}$ & $\sim 700$  \\
\rule[-1ex]{0pt}{3.5ex}  $z\sim 4$ & $>3\times 10^{12}$ & $>300$ & $>10^{11}$ & $\sim 30$  \\
\hline
\end{tabular}
\end{center}
\caption{Expected number of detections and physical sensitivity limits with {\it PRIMA} in the PRIDES survey over $1.6\,{\rm deg^2}$ over the COSMOS field.\label{tab:numbers}} 
\end{table} 

{\it PRIMA} is crucial to pursue the above science goals: neither JWST nor ALMA can observe the rest-frame mid-IR (silicate features to IR SED peak) of $z > 3$ galaxies. Note that only some observations by JWST/MIRI of PAHs exist for AGN at $z\sim0.5$\cite{young23} and a lensed galaxy at $z\sim4.2$ \cite{spilker23}. {\it PRIMA} will extend these studies to more typical galaxies at are not lensed.
A large field of view and fast mapping speed is crucial for obtaining a large survey area to {\it (i)} cover the Universe’s large- scale structure and diversity of environments, {\it (ii)} probe a variety of galaxy properties (from main-sequence to starburst galaxies), and {\it (iii)} pursue a true flux-limited blind survey to mitigate various biases introduced by targeted surveys (e.g., UV selection bias that is currently in place in targeted studies with ALMA). Table~\ref{tab:numbers} lists the expected number of detections and limits in IR luminosity, star formation rates (SFR), and stellar masses of the PRIDES survey. These numbers are derived from the COSMOS2020 catalog\cite{weaver22}, requiring a detection in the bluest PRIMAger bands\footnote{Note that due to the rise of the mid-IR SED, a detection in the blue PRIMAger band will generally result in a detection in the redder bands.} given its current expected limiting sensitivities (see Ciesla et al. {\it in prep.}, this volume). Specifically, we used the average far-IR SED of observed starburst galaxies\cite{kirkpatrick12} to translate the PRIMAger sensitivity limits in the mid-IR into a total IR luminosity. This we convert to a SFR using common relations\cite{kennicutt98} and use the main-sequence relation between stellar mass and SFR\cite{schreiber15,khusanova21} to derive a stellar mass. We then count the number of sources in the COSMOS2020 catalog (limited to the PRIDES area) that satisfy these limits in four redshift bins with $\Delta z = 1$ (considering only sources with their $1\sigma$ redshift probability fully within these redshift windows). We only consider robust detections in the COSMOS2020 catalog ({\it i.e.} with stellar mass measurement, secure photometric redshifts, and not flagged to have corrupted photometry), which may result in a lower limit of number counts.

The left panel in Figure~\ref{fig:sed} shows the expected depth and wavelength coverage of PRIDES. Shown are four different model SEDs of galaxies at $z=1-4$ normalized to different IR luminosities $\LIR$. The corresponding stellar masses are derived assuming a common relation between $\LIR$ and SFR\cite{kennicutt98}. At the suggested depth of PRIDES of $70\,{\rm \mu Jy}$, we expect to detect the mid-IR continuum of galaxies at $z=3$ at a stellar mass limit of $3\times 10^{10}$, corresponding to $\LIR = 10^{12}\,{\rm L_\odot}$. In the case of a dust-obscured AGN, the mid-IR emission is boosted, which would result in lower limiting luminosities.

PRIDES will be covering the COSMOS field\cite{scoville07} (see Section~\ref{sec:layout}), which provides a wealth of ancillary data at comparable or deeper depth (left panel of Figure~\ref{fig:sed}). {\it PRIMA} will be synergistic with these data, providing the crucial missing puzzle piece at mid-IR wavelengths.
COSMOS has been covered by multiple spectroscopic surveys providing more than 100,000 spectroscopic redshifts (and emission line measurement for a fraction)\cite{lefevre13,hasinger18,khostovan25}.
The ancillary data also include deep observations with {\it Spitzer} at $\sim0.3\,{\rm \mu Jy}$ at $3.6\,{\rm \mu m}$ and $4.5\,{\rm \mu m}$ as well as $\sim 3\,{\rm \mu Jy}$ at $5.7\,{\rm \mu m}$ and $8\,{\rm \mu m}$. This is significantly deeper than the PRIDES depth and therefore provides a connection between {\it PRIMA}'s bluest bands and JWST's reddest bands.
In addition, the PRIDES field is imaged by HST/ACS in F814W\cite{koekemoer07} as well as JWST/NIRCam observations (at F115W, F150W, F277W, and F444W) and JWST/MIRI observations (at F770W and F1800W on a fraction of the area) from the large JWST programs {\it COSMOS-Web} (PI: Kartaltepe \& Casey)\cite{casey23} and PRIMER (PI: Dunlop). These surveys provide rest-frame UV and optical observations across the PRIDES field\footnote{{\it COSMOS-Web} covers $0.6\,{\rm deg^2}$ of the total $1.6\,{\rm deg^2}$ area of PRIDES.}. The sensitivity flux densities of these surveys are $<10^{-6}\,{\rm \mu Jy}$ (significantly deeper than PRIDES) and are therefore not indicated in Figure~\ref{fig:sed} for clarity.
In addition, PRIDES is synergetic with sub-mm observations from the CHAMPS (Faisst et al., {\it in prep.}) and the (extended) MORA \cite{casey21,long24} ALMA programs at $1.2\,{\rm mm}$ and $2\,{\rm mm}$ as well as single-dish surveys including the NIKA2 Cosmological Legacy Survey at $1.2\,{\rm mm}$ using the $30\,{\rm m}$ IRAM radio telescope\cite{bing23} and the SCUBA-2 survey using JCMT at $850\,{\rm \mu m}$\cite{simpson19}. Their data will detect the far-IR SED {\it redward} of the peak ($>100\,{\rm \mu m}$ rest-frame) for any detected PRIDES galaxy (see left panel in Figure~\ref{fig:sed}).
The sensitivity limits of retired facilities\cite{Jin18,frayer09} such as {\it Spitzer}/MIPS and {\it Herschel}/PACS and SPIRE are also indicated on the left panel of Figure~\ref{fig:sed}. However, except the {\it Spitzer} $24\,{\rm \mu m}$ data, these do not add significant constraints on the mid- to far-IR SEDs as mentioned earlier other than from stacking analyses. Note that although the {\it Spitzer} $24\,{\rm \mu m}$ sensitivity is comparable to that in the bluest band of PRIMAger, it only provides one photometric point and is therefore not significantly constraining redshifts or total IR luminosities.
The PRIDES field will also be targeted by the {\it Euclid} and later {\it Roman} space telescopes providing rest-frame optical spectroscopy. In addition, the new SPHEREx telescope\cite{crill20} (to be launched in spring 2025) will perform an all-sky survey using LVFs (similar to {\it PRIMA}) in $102$ filters from $0.75\,{\rm \mu m}$ to $5\,{\rm \mu m}$ at a depth of $\sim 30\,{\rm \mu Jy}$ at $5\sigma$. We note that this is deeper compared to the PRIDES depth and therefore provides a seemless connection to rest-frame optical spectroscopy for the brightest emission line galaxies identified in PRIDES.

The right panel in Figure~\ref{fig:sed} shows the simulated {\it PRIMA} mid-IR photometry in PRIDES of two SEDs derived from a library of AGN and non-AGN templates\cite{kirkpatrick12,kirkpatrick15} normalized to $\LIR = 10^{12}\,L_\odot$ at $z=3$.
The peak at $\sim 7.7\,{\rm \mu m}$ in the SED without AGN contribution (purple squares) indicates a strong PAH spectral band. This PAH emission is caused by small dust grains in the ISM and is related to SFR and metallicity\cite{whitcomb24,madden06,gordon08,egorov23}.
The AGN shown in this example has a bolometric luminosity of $\sim 10^{45}\,{\rm erg\,s^{-1}}$, derived using the relations from Spinoglio et al. (2024)\cite{spinoglio24}. This luminosity is about one order of magnitude below the knee of the AGN bolometric luminosity function at $z\sim3$\cite{shen20} and furthermore is representing the bulk of detected Type 1 and 2 AGN in the COSMOS field\cite{lusso12}.
This shows that PRIDES will be able to detect PAH emission out to $z=3$ regularly in such galaxies and that it covers a representative population of AGN in the COSMOS field at these wavelengths, which has not been able to be done in the past before {\it PRIMA}. The large survey area of PRIDES will enable this measurement in hundreds of $z=3$ galaxies (and thousands at $z<3$, Table~\ref{tab:numbers}) at different stellar masses and SFR, with additional information on metallicity, resulting in a breakthrough of the study of small dust-grain properties out to high redshifts.
On the other hand, the yellow circles show an SED with $80\%$ contribution to the total mid-IR flux from an AGN. The abundance of small grain is reduced due to the hard radiation and heating from the central AGN (as shown in spatial resolved studies of local galaxies\cite{lai23}), resulting in a reduction of the PAH emission.
This simulation demonstrates that {\it PRIMA} is an excellent facility thanks to its sensitivity and multi-band resolution to measure the broad PAH emission and identify dust-obscured AGN disintegrating and removing dust grains.
On the other hand, there is the possibility of significant shielding of the AGN torus in a very dust obscured system, which would result in the preservation of the PAH emission\cite{alonsoherrero14}. In that case, the Silicate feature at $9.7\,{\rm \mu m}$ (which tends to be in absorption in dust-obscured AGN\cite{garciabernete22}), which loosely correlates with the amount of obscuration, could be used as additional tracer of dust-obscure AGN activity\cite{vignali11, gonzalezmartin13, lacaria19}.

We note that the confusion will be significant given the large pixel sizes of {\it PRIMA} (see left panel of Figure~\ref{fig:sed}). However, using ancillary data from ground-based telescopes, {\it Spitzer} and JWST as well as making use of the bluer bands of {\it PRIMA}, confusion can be readily reduced. For this end, forced photometry methods including the spatial (and structural) priors from these bluer bands can be used (see Section~\ref{sec:layout} for more details).

\begin{figure*}
\begin{center}
\includegraphics[width=0.65\textwidth]{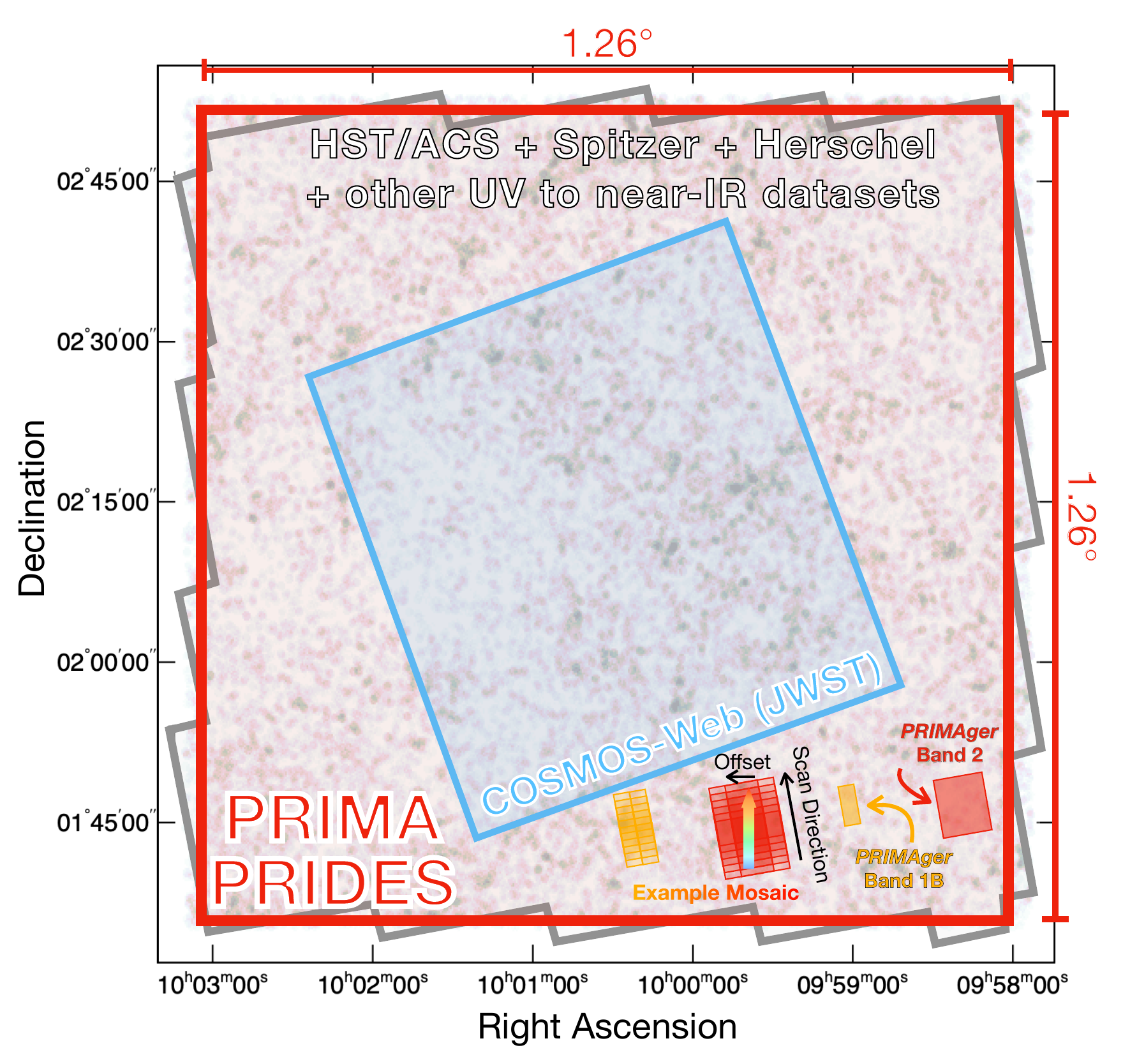}\vspace{-5mm}
\end{center}
\caption 
{\label{fig:layout}
Layout of the $1.6\,{\rm deg^2}$ PRIDES observations (red) on the COSMOS field\cite{scoville07}. The {\it COSMOS-Web} JWST survey\cite{casey23} is indicated in blue. The gray outline marks the coverage of the COSMOS HST/ACS observations\cite{koekemoer07}. COSMOS is observed with a wealth of other facilities from UV to sub-mm. The background cloud shows the large-scale structure of galaxies at $3 < z < 4$ (from the \texttt{Galacticus} simulation\cite{benson12}). The PRIDES area is optimal to cover various environmental properties including voids and overdensities. PRIMAger footprints for Band 1 (orange) and Band 2 (red) are shown on the lower right.} 
\end{figure*} 

\subsection{PRIDES -- Survey Layout and Technical Details}\label{sec:layout}

The PRIDES program would survey a total area of $1.6\,{\rm deg^2}$, covering the HST/ACS footprint on the COSMOS field (Figure~\ref{fig:layout}) with the PRIMAger instrument to observed typical (main-sequence) galaxies mainly out to $z\sim4$ with total IR luminosities of $>10^{12}\,{\rm L_\odot}$ (corresponding of a SFR of $>100\,{\rm M_\odot\,yr^{-1}}$ or a stellar mass of $>3\times 10^{10}\,{\rm M_\odot}$) in Band 1 ($25-80\,{\rm \mu m}$). Note that Band 2, including polarimetry, at 96, 126, 172, and $235\,{\rm \mu m}$ is observed simultaneously for free, but we expect a significant confusion (which can be remedied somewhat by smart flux measurements, see below) and therefore most of the high-redshift science will focus on Band 1. As shown in the left panel of Figure~\ref{fig:sed}, observed $50 - 100\,{\rm \mu m}$ fluxes of typical $z=1-4$ main-sequence galaxies are on the order of $70\,{\rm \mu Jy}$ based on recent studies with ALMA and JWST\cite{faisst20,talia21,gruppioni20,mckinney23}. Assuming an average sensitivity of $250\,{\rm \mu Jy}$ ($5\sigma$) in 10 hours over $1\,{\rm deg^2}$, PRIDES would need a total of $\sim200\,{\rm hours}$ (including an efficiency of $65\%$) to survey an area of $1.6\,{\rm deg^2}$ to a depth of $70\,{\rm \mu Jy}$ ensuring a $5\sigma$ continuum detection of these galaxies.
We note that such a time request would not be outrageous. Current large survey with ALMA and JWST are on the order of $100-200\,{\rm hours}$, but on smaller areas (e.g., {\it COSMOS-Web} covers $0.6\,{\rm deg^2}$ and CHAMPS covers $0.18\,{\rm deg^2}$). It is planned that $75\%$ over a 5-year nominal mission timeline is devoted to GO science. PRIDES would use $8-10\,{\rm days}$ of this time and results in a significant scientific return.

To cover the requested area, the observations will have to be carried out in mapping mode (scanning in wavelength direction and spatial offsets in perpendicular direction as visualized in Figure~\ref{fig:layout}). The orientation angle would not be constrained in order to maximize scheduling of the observations. Ideally, the observations should be taken at the same orientation angle to maximize consistency over the field and minimize data reduction efforts (e.g., PSF orientation, deblending, etc.). This will achieve a continuous coverage in PRIMAger Band 1A (which covers the smallest FoV) and since the other bands overlap their resulting depth will be slightly increased.

\begin{figure*}
\begin{center}
\includegraphics[width=0.8\textwidth]{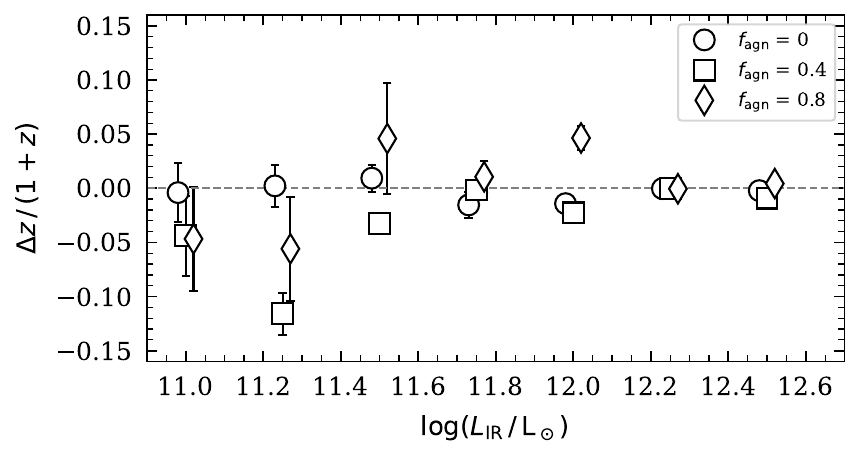}\vspace{-5mm}
\end{center}
\caption 
{
Expected redshift accuracy as a function of total IR luminosity and for different contributions of AGN. The redshifts are mainly determined from the PAH spectral feature. This leads to a worse redshift accuracy for enhanced AGN contribution (which tends to weaken the PAH emission).
\label{fig:redshiftsim}}
\end{figure*}

The planned depth of PRIDES is deeper than wide-field surveys planned with {\it PRIMA} as part of the PI science cases. PRIDES therefore optimally complements such surveys following a ``wedding cake'' strategy. However, the downside is that this requested depth is at the confusion limit in the PRIMAger Band 1\cite{bethermin24,donnellan24}. PRIDES will use efficient and robust deblending methods making use of deep near-IR and IR imaging ancillary data on the COSMOS field for flux and positional priors to push {\it PRIMA}’s effective sensitivity limits to the maximum. These techniques have been used commonly to de-confuse {\it Spitzer} and {\it Herschel} as well as ground-based data\cite{bethermin15,weaver22,faisst22,liu18,Jin18}. Recently, such deblending algorithms have been tested for {\it PRIMA}\cite{donnellan24}. Furthermore, these techniques will also be used to measure photometry from data of the new SPHEREx space telescope\cite{crill20} which will be at similar pixel scales as {\it PRIMA}.
Common software to do this may include \texttt{Galight}\cite{ding21} or \texttt{Tractor}\cite{lang16,weaver23}, which both are based on a forward modeling approach of source prior positions and shapes to account for blending of datasets with different PSF sizes, pixel scales, and sensitivities.
For deblending, PRIDES would make use the existing deep UltraVISTA $2.2\,{\rm \mu m}$ Ks-band ($\sim25.2\,{\rm AB}$, $3\sigma$) and {\it Spitzer} $3.6\,{\rm \mu m}$ and $4.5\,{\rm \mu m}$ ($\sim26.3\,{\rm AB}$, $3\sigma$) imaging as well as the blue PRIMAger bands for positional priors. In the {\it COSMOS-Web} footprint (covering 40\% of the PRIDES footprint), we will in addition make use of the deep JWST/NIRCam F227W and F444W imaging ($\sim28.5\,{\rm AB}$, $5\sigma$). Band 2 will be observed simultaneously for free (including polarimetry), but these observations will suffer higher confusion (see left panel in Figure~\ref{fig:sed}). Deblending may be possible using the bluer wavelength of Band 2 and Band 1 specifically. Moreover, depending on the observation strategy ({\it i.e.}, number of frames per exposure and number of scans across the field) and data reduction (e.g., reducing half of the frames), the impact of confusion could be reduced in Band 2.

Figure~\ref{fig:redshiftsim} shows the expected redshift accuracy achieved by PRIDES over the COSMOS field for $z=1-4$. The simulation, applying realistic PRIDES sensitivity limits, was done for SED templates with different contribution from AGN\cite{kirkpatrick12} (similar to those shown on the right panel of Figure~\ref{fig:sed}) and varying total IR luminosities. The accuracy, expressed as $\Delta z\,/\,(1+z)$, is generally $<0.03$ at $>4\times10^{11}\,{\rm L_\odot}$. Because the main spectral feature are the PAH emission line bands, the redshift accuracy is lower for AGN dominated systems, as the presence of an AGN generally weaken the PAH emission.

For efficiency reasons, spectroscopy for such a program would not be requested over such a large area as the broad PAH spectral features and the mid--IR slope can be measured well with model SED fitting at the resolution of the photometry channels (see right panel in Figure~\ref{fig:sed}). Redshifts of dust-obscured galaxies can be confirmed directly using these spectral features using template fitting (resulting in redshifts accuracies comparable with those from narrow-band emission line surveys, see Figure~\ref{fig:redshiftsim}). If no such features are present, the additional continuum point will greatly enhance the redshift accuracy compared to one single sub-mm observations (in the case of optical dark galaxies).

\section{Conclusions}\label{sec:conclusions}

 Currently, there does not exist a facility with the sensitivity and mapping speed of {\it PRIMA} that covers the wavelength gap from $24-235\,{\rm \mu m}$ between the JWST and ALMA observatories. {\it PRIMA} will therefore provide a crucial turning point in the study of AGN, black hole growth, and dust production of galaxies in the early universe out to (at least) $z=4$. In addition, {\it PRIMA} enables a unique synergy with other current and future facilities. To mention a few related to high-redshift studies:
 \begin{itemize}
     \item  It provides the necessary wavelength coverage to understand the SED shape and the characteristics (including distances) of elusive dust-obscured galaxies found by JWST and ALMA.
    \item  The multi-band resolution in the mid-IR range enables the measurement of PAH emission and silicate absorption, which can be linked to measurements from the optical lines (from JWST or {\it Euclid}) such as metallicity or ionization parameters to understand dust production mechanisms and the dependence of abundance of dust to ISM conditions.
    \item The lack of PAH emission and steep rest-frame mid-IR slopes are a way to efficiently select dust-obscured and Compton-thick AGN, which are missed by other observatories working at UV/optical wavelengths and X-ray frequencies.
 \end{itemize}

A survey such as PRIDE over an area of $1.6\,{\rm deg^2}$ (making use of {\it PRIMA}'s sensitivity and mapping speed) is feasible within less than $300\,{\rm hours}$ of total observing time and would provide enough area to carry out the above studies with a statistical sample of $10,000$s of galaxies out to $z=4$. Carrying out such a survey on the COSMOS field has the advantage of adding a wealth of ancillary data. COSMOS will also be observed in the future by {\it Euclid}, SPHEREx, and {\it Roman} thus increasing the importance of {\it PRIMA} in the next decade. The legacy of such a survey would be tremendous and would enable science beyond the science goals mentioned above. A clever survey layout strategy (``stepping'' will be necessary due to the LVF-nature of the exposures) could also be used to enable transient searches at mid-IR wavelengths at higher redshifts, similar to studies with WISE that target variable AGN at $z<1$ and other transient phenomena\cite{prakash19,faisst19}.

\subsection* {Code, Data, and Materials Availability} 
The {\it PRIMA} sensitivity data can be obtained from the respective studies published in the same special volume as this work.

\subsection* {Disclosures}
The authors declare there are no financial interests, commercial affiliations, or other potential conflicts of interest that have influenced the objectivity of this research or the writing of this paper.

\subsection* {Acknowledgments}
The authors declare there are no financial interests, commercial affiliations, or other potential conflicts of interest that have influenced the objectivity of this research or the writing of this paper


\bibliography{bibli}   
\bibliographystyle{spiejour}   


\vspace{2ex}\noindent\textbf{Andreas Faisst} is an assistant research scientist at IPAC, which is part of the California Institute of Technology (Caltech) in Pasadena, USA. He obtained his PhD degree at the ETH Zurich in 2015. Since 2013, he led more than 16 papers and contributed to more than 115 publications. He is leading several large collaboration such as the {\it ALPINE} and {\it CHAMPS} large ALMA programs and several programs with the JWST. His research interests include physics in the Epoch of Reionization, early galaxy evolution, the dust properties of galaxies, AGN, and quiescent galaxies at cosmic noon.
\vspace{1ex}
\noindent Biographies and photographs of the other authors are not available.

\listoffigures
\listoftables

\end{spacing}
\end{document}